\documentclass[12pt]{article}
\global\parskip 6pt
\newcommand{\bea}{\begin{eqnarray}}
\newcommand{\eea}{\end{eqnarray}}
\newcommand{\non}{\nonumber}

\begin{document}
\begin{titlepage}
\vspace*{.5cm}
\begin{center}
{\Large{{\bf Stability of Topological Black Holes}}} \\[.5ex]
\vspace*{1.5cm} Danny Birmingham\footnote{Email: dbirmingham@pacific.edu}\\
\vspace{.1cm}
{\em Department of Physics,\\
University of the Pacific,\\
Stockton, CA 95211\\
USA}\\
\vspace*{1cm} Susan Mokhtari\footnote{Email:
susan@science.csustan.edu}\\
\vspace{.1cm} {\em Department of Physics,\\
California State University Stanislaus,\\
Turlock, CA 95380,\\
USA}
\begin{abstract}
\noindent {We explore the
classical stability of topological black holes
in $d$-dimensional anti-de Sitter spacetime, where the horizon
is an Einstein manifold of negative curvature.
According to the gauge invariant formalism of
Ishibashi and Kodama, gravitational
perturbations are classified as being of scalar, vector, or tensor type,
depending on their transformation properties with respect
to the horizon manifold.
For the massless black hole, we show that the perturbation equations
for all modes can be reduced to a
simple scalar field equation. This equation is exactly solvable in
terms of hypergeometric functions, thus allowing an exact analytic
determination of potential gravitational instabilities.
We establish a necessary and sufficient
condition for stability,
in terms of the eigenvalues $\lambda$ of the Lichnerowicz operator
on the horizon manifold, namely $\lambda \geq -4(d-2)$.
For the case of negative mass black holes,
we show that a sufficient condition for stability
is given by $\lambda \geq -2(d-3)$.}
\\
\end{abstract}
\vspace*{.25cm} September 2007
\end{center}
\end{titlepage}

\section{Introduction}

Black holes in anti-de Sitter space have been the subject of much recent attention, particularly
in connection with the proposed correspondence between anti-de Sitter gravity and boundary conformal
field theory (AdS/CFT) \cite{Maldacena}-\cite{Witten}.
A class of static black holes solutions to $d$-dimensional anti-de Sitter gravity
was constructed in \cite{Birtopbh}, with the special property that the
horizon $M^{d-2}$ is a $(d-2)$-dimensional compact Einstein manifold of
positive, zero, or negative curvature. Furthermore, for the case of negative curvature horizon, there is a class of black holes
for which the mass parameter $M$ can assume both negative and zero values,
$M_{\mathrm{ crit}} \leq M \leq 0$.
While topological black holes allow one to study the boundary conformal
field theory on spaces of the form $S^{1}\times M^{d-2}$ \cite{Birtopbh}, they  are also interesting structures in their own right.
It is of particular interest to explore their classical stability properties.

The response of a black hole metric to small perturbations has been
the subject of investigation for many years.
By analyzing the perturbation equations
subject to certain boundary conditions, one can gain valuable insight into the
structure of the black hole.  In particular, one can address the question of the classical stability of the black hole.
For the Schwarzschild black hole in four
dimensions, this boundary value problem was analyzed quite some time ago.
It was shown that metric perturbations could be described by either a scalar mode (the Zerilli mode \cite{Z1,Z2})
or a vector mode (the Regge-Wheeler mode \cite{RW}). Using the Zerilli and Regge-Wheeler equations, the stability
was then established in \cite{Vish}-\cite{Wald}.
Remarkably, a similar analysis for the higher-dimensional
Schwarzschild black hole \cite{Tangherlini} was lacking until
quite recently. In \cite{Gibbons}, it was shown that an additional tensor mode is present in
dimensions greater than four. Following this, a complete
gauge invariant formalism was developed by Ishibashi and Kodama \cite{Kodama1},
where the equations describing all gravitational perturbations of all higher-dimensional
static black holes have been presented. This powerful formalism identifies three basic types of
gravitational master field, depending on how the field transforms with respect to the horizon manifold.
One has scalar and vector modes, and an additional tensor mode in dimensions greater than four.
Moreover, the equations for these master fields have a standard form as a Schr\"{o}dinger-type second order ordinary differential
equation with a potential.

The Ishibashi-Kodama formalism has been used successfully to establish the stability of a wide array
of charged and uncharged black holes in asymptotically flat, de Sitter,
and anti-de Sitter space \cite{Kodama2}-\cite{Kodama5}.
Generalizations of this formalism to include the effects of
rotation have also been
considered \cite{Reall}.
In particular, the stability of the asymptotically flat Schwarzschild black hole was firmly established in
all dimensions, following earlier work in \cite{Gibbons}. The key technique used in
\cite{Kodama2}-\cite{Kodama5} is a so-called $S$-deformation of the potentials appearing in the master equations.
This technique allows one to establish stability based on positivity of the corresponding deformed potentials.
However, apart from the asymptotically flat Schwarzschild black hole, there is no other class of black holes
for which stability has been established in all dimensions.

Our goal here is to investigate the stability properties of
topological black holes with negative curvature horizon, for mass
parameter in the range $M_{\mathrm{crit}} \leq M \leq 0$. We first
consider the massless topological black hole, and show that the
master field equations for all modes can be solved explicitly in all
dimensions \cite{BM}. The solution can be written in terms of
hypergeometric functions, and by imposing appropriate boundary
conditions, we can analyze the stability question in full detail.
Using this explicit solution, we are led directly to a necessary and
sufficient condition for stability of the massless black hole. This
calculation completes and extends earlier work in
\cite{Gibbons,Kodama3}. We show that stability is determined solely
by the spectrum of the Lichnerowicz operator on the horizon
manifold. In particular, we conclude that massless topological black
holes with either constant curvature (hyperbolic) horizons, or
Einstein-K\"{a}hler horizons, are stable in all dimensions.
Following this, we consider the case of negative mass black holes.
Although the explicit solution of the master equations is not
available, we are nevertheless able to establish a sufficient
condition for stability. Using positivity of the gravitational
potentials, as well as the $S$-deformation technique, allows us to
derive a sufficient condition for stability in all dimensions. As an
example, we show that negative mass black holes with
Einstein-K\"{a}hler horizon are stable in all dimensions.
It should be pointed out that, in most cases, the boundary conditions are
dictated by the requirement of normalizability of the perturbation.
As we shall see, we must then impose Dirichlet boundary conditions
at both the horizon and infinity. However, there is additional freedom
in the choice  of boundary conditions for certain perturbations
in dimensions four, five, and six,
as pointed out in \cite{Wald2}.

The plan of this paper is as follows. In section 2, we recall the
essential features of topological black holes in anti-de Sitter
space. In particular, we identify the massless and negative mass
black holes with negative curvature horizon, which are the main
focus of interest. In section 3, we present the basic equations in
the Ishibashi-Kodama gauge invariant formalism for gravitational
perturbations. In section 4, we demonstrate the unified form which
these equations take for the case of the massless black hole. The
explicit solution of the master equations is given and a necessary
and sufficient condition for stability is derived. Section 5 deals
with negative mass black holes, and we use positivity properties of
the gravitational potentials to establish a sufficient condition for
stability. We conclude in section 6 with a brief discussion of our
results, and also highlight the issue of boundary conditions in
dimensions four, five, and six.

\section{Topological Black Holes in anti-de Sitter Space}
In $d$-dimensional anti-de Sitter space, there is a class of topological black hole solutions
to Einstein's equations which has the property that
the horizon $M^{d-2}$ is a $(d-2)$-dimensional compact Einstein space of positive, zero
or negative curvature $k$ \cite{Birtopbh}. Topological black holes solutions in four dimensions were first constructed in
\cite{Lemos}-\cite{Brill}.
The line element of the topological black hole is given by \cite{Birtopbh}
\bea
ds^{2} = - f(r)\; dt^{2} + f^{-1}(r)\;dr^{2} +
r^{2}h_{ij}(x)\;dx^{i}dx^{j},
\label{topbh}
\eea
where
\bea
f(r) = \left(k -
\frac{\omega_{d}M}{r^{d-3}} +\frac{r^{2}}{l^{2}}\right),
\eea
and
\bea
\omega_{d} = \frac{16\pi G}{(d-2)\mathrm{Vol}(M^{d-2})}.
\eea
The parameter $k$ can take the values $k=1, 0, -1$.
The volume of the horizon is denoted by $\mathrm{Vol}(M^{d-2}) = \int d^{d-2}x\;\sqrt{h}$.
The parameter $l$, with dimensions of length, is related to the cosmological constant $\Lambda$ by
$\Lambda = -(d-1)(d-2)/2l^{2}$, and $\omega_{d}$ is inserted so that $M$ has dimensions of inverse length.

It is straightforward to check that the metric (\ref{topbh})
satisfies Einstein's equations with negative cosmological constant,
namely \bea R_{\mu\nu} = -\frac{(d-1)}{l^{2}}g_{\mu\nu}, \eea
provided that the horizon is an Einstein space
\bea R_{ij}(h) = k(d-3)h_{ij}. \eea Our interest here is in the
negative curvature case with $k=-1$. An interesting subclass of
black holes is then obtained by taking the horizon to be a manifold
of constant curvature, i.e., a hyperbolic manifold. In this case,
$M^{d-2} = H^{d-2}/\Gamma$, where $H^{d-2}$ is hyperbolic space and
$\Gamma$ is a suitable discrete subgroup of the isometry group of
$H^{d-2}$.

The mass parameter $M$ can be expressed in terms of the location of the horizon $r_{+}$, as
\bea
M = \frac{r_{+}^{d-3}}{\omega_{d}}\left(-1 + \frac{r_{+}^{2}}{l^{2}}\right).
\label{Mass}
\eea
Furthermore, the inverse Hawking temperature is given by \cite{Birtopbh}
\bea
\beta = \frac{4\pi l^{2}r_{+}}{(d-1)r_{+}^{2} - (d-3)l^{2}}.
\eea
A very special feature which is present in the
case of negative curvature horizon, is that the parameter
$M$ can assume negative values, as first discussed in
\cite{Vanzo,Mannneg1,Brill,Mannneg2}.
The requirement of positivity of temperature
enforces an inequality on the value of $r_{+}$, namely that
$r_{+} > r_{\mathrm{crit}}$, where
\bea
r_{\mathrm{crit}} = \left(\frac{d-3}{d-1}\right)^{1/2}l.
\eea
The corresponding value of $M$ is then given by (\ref{Mass}),
\bea
M_{\mathrm{crit}} = -\left(\frac{2}{d-1}\right)\left(\frac{d-3}{d-1}\right)^{(d-3)/2}\frac{l^{d-3}}{\omega_{d}}.
\eea
Thus, when $k=-1$, there is a class of black holes with  mass
parameter $M$ in the range $M_{\mathrm{crit}} \leq M \leq 0$.
For $M=0$, we note that the event horizon occurs at $r_{+} = l$, while
for $M_{\mathrm{crit}} \leq M < 0$, we have $r_{+} < l$.
It should be noted that the $M = M_{\mathrm{crit}}$ solution has a
degenerate horizon at $r = r_{\mathrm{crit}}$ with
$f(r_{\mathrm {crit}}) = f^{\prime}(r_{\mathrm{crit}}) = 0$.
Although these extremal solutions do not strictly have an interpretation as
black holes \cite{Vanzo, Brill}, we can
still incorporate them into the stability analysis that follows.

These topological black holes are interesting structures in their
own right, and our goal here is to investigate their classical stability properties.
However, they also assume an importance within the context of the AdS/CFT correspondence.
In particular, they allow us to study
the dual conformal field theory on spaces of the form
$S^{1} \times M^{d-2}$, where
$M^{d-2}$ is an Einstein space of
positive, zero, or negative curvature \cite{Birtopbh, Emparan}.

\section{Gravitational Perturbations}
In order to check for the existence of unstable gravitational perturbations of a black hole, we first need to
obtain the relevant equations which describe these perturbations. In the four-dimensional
asymptotically flat case, this was achieved quite some time ago, resulting in the Zerilli equation \cite{Z1,Z2}
and the Regge-Wheeler equation \cite{RW}. These equations were generalized to the anti-de Sitter case in \cite{Cardoso2,
Cardoso3}.
However, a general analysis of gravitational perturbations in higher dimensions was
presented only recently by Ishibashi and Kodama \cite{Kodama1}-\cite{Kodama5}.
The formalism developed by Ishibashi and Kodama is both powerful and elegant, and is based on
the introduction of gauge invariant variables. These gauge invariant combinations are
then described by master fields $\Phi$. In general,
there are three types of gravitational perturbation; the scalar mode which is the analogue of the Zerilli mode
in higher dimensions, the vector mode which is the analogue of the Regge-Wheeler mode, and an additional
tensor mode which is present in dimensions greater than four \cite{Gibbons}. As shown in \cite{Kodama1}, each perturbation
is simply described in terms of a master field $\Phi$ which satisfies a
Schr\"{o}dinger-type
second order  ordinary differential equation.

The Ishibashi-Kodama  equations have been used to successfully establish the stability
of asymptotically flat Schwarzschild black holes in all dimensions \cite{Kodama2}, following an earlier
analysis in \cite{Gibbons}.
However, for many other higher-dimensional black holes, the stability question is still an open issue,
with only partial results available.
Our aim here is to find several new examples where precise conditions for stability can be established in all dimensions.
The stability of topological black holes with respect to scalar field couplings
has been analyzed in \cite{Chan, Wang}.

To begin, we write the master field as
\bea
\Phi(t,r,x^{i}) = \Phi(r)Y(x^{i})e^{\omega t}.
\label{ansatz}
\eea
The type of perturbation then depends on whether $Y$ transforms as a scalar, vector, or tensor
with respect to the horizon manifold $M^{d-2}$. In all cases, however, the
master equation takes the simple form
\bea
\left[-\left(f\frac{d}{dr}\right)^{2} + V\right]\Phi(r) = - \omega^{2}\Phi(r),
\label{master}
\eea
where the structure of the potential $V$ depends on the gravitational mode
under consideration. For the scalar mode, we have
\bea
V_{\mathrm{S}}(r) = \frac{fU(r)}{16r^{2}H^{2}},
\label{Scalar}
\eea
where
\bea
x &=& \frac{\omega_{d} M}{r^{d-3}},\;\;\mu = k_{\mathrm{S}}^{2} + (d-2),\non\\
H &=& \mu + \frac{1}{2}(d-2)(d-1)x.
\eea
In this case, $Y$ transforms as a scalar, and is an eigenfunction of the scalar Laplacian
on the horizon manifold,
$\nabla^{2}Y = -k_{\mathrm{S}}^{2}Y$.
The function $U(r)$ is given by
\bea
U(r) &=& [(d-2)^{3}d(d-1)^{2}x^{2} - 12(d-2)^{2}(d-1)(d-4)\mu x
+ 4(d-4)(d-6)\mu^{2}]\frac{r^{2}}{l^{2}}\non\\
&+& (d-2)^{4}(d-1)^{2}x^{3} + (d-2)(d-1)[4(2(d-2)^{2} - 3 (d-2) + 4)\mu\non\\
&-& (d-2)(d-4)(d-6)(d-1)]x^{2}
- 12(d-2)[(d-6)\mu \non\\
&-& (d-2)(d-1)(d-4)]\mu x
+ 16\mu^{3} -4(d-2)d\mu^{2}.
\eea
The vector mode is described by the potential
\bea
V_{\mathrm{V}}(r) = \frac{f}{r^{2}}\left[k_{\mathrm{V}}^{2} - 1 -
\frac{(d-2)(d-4)}{4} + \frac{(d-2)(d-4)}{4}\frac{r^{2}}{l^{2}} - \frac{3 (d-2)^{2}\omega_{d}M}{4r^{d-3}}\right],
\label{Vector}
\eea
where $\nabla^{2} Y  = -k_{\mathrm{V}}^{2}Y$.
Finally, the tensor mode in dimension $d>4$ has the potential
\bea
V_{\mathrm{T}}(r) = \frac{f}{r^{2}}\left[\lambda +2(d-3) -
\frac{(d-2)(d-4)}{4} + \frac{d(d-2)}{4}\frac{r^{2}}{l^{2}} + \frac{ (d-2)^{2}\omega_{d}M}{4r^{d-3}}\right],
\label{Tensor}
\eea
where $\lambda$ is the eigenvalue of the Lichnerowicz operator on the horizon.

In order to investigate the stability properties of the black hole,
it is useful to re-cast Eq. (\ref{master})
in the form
\bea
A\Phi = -\omega^{2}\Phi,
\label{Aoperator1}
\eea
where
\bea
A = -\frac{d^{2}}{dr_{*}^{2}} + V(r),
\label{Aoperator2}
\eea
and the tortoise coordinate $r_{*}$ is defined by $dr_{*} = \frac{dr}{f}$.
Our task is to solve this equation subject to appropriate boundary conditions.
In particular, unstable
modes correspond to normalizable negative energy ($\omega > 0$) states
of the Schr\"{o}dinger operator $A$.
In order to ensure normalizability, in the sense that \cite{Gibbons, Kodama2},
\bea
1 = \int dr_{*}\;\Phi^{*}\Phi,
\label{norm}
\eea
we must impose boundary conditions both at the horizon
and infinity.
Near the horizon, normalizability demands that we impose a Dirichlet
boundary condition $\Phi \rightarrow 0$ on the perturbation
\cite {Gibbons, Kodama2, KayWald}.
For large $r$, we see that the perturbation must behave as
$\Phi \sim r^{\alpha/2}$ as $r \rightarrow \infty$, with $\alpha < 1$.
For dimensions $d\geq 7$, the latter requirement is only satisfied by imposing Dirichlet boundary conditions
on the perturbation. However, for
certain perturbations in dimensions four,
five, and six, there is additional freedom
in the choice of boundary conditions
at infinity. In fact, as shown in \cite{Wald2} for the
case of pure anti-de Sitter space,
there is a one-parameter family of self-adjoint extensions of the
operator $A$ in these cases.
Each of these self-adjoint extensions comes equipped with a
choice of boundary conditions at infinity.
Since the asymptotic form of the metric for topological black holes (\ref{topbh})
is akin to the pure anti-de Sitter case, we
observe a similar freedom in the choice of boundary conditions
for these these perturbations.
In the following, we will study the stability properties in
these special cases with respect to a choice of Dirichlet
boundary conditions at infinity.

\section{Stability of the Massless Black Hole}
The form of the potentials (\ref{Scalar}), (\ref{Vector}), and (\ref{Tensor})
simplifies considerably for the massless topological black hole $M=0$.
The scalar potential is given by
\bea
V_{\mathrm{S}} = \frac{f}{r^{2}}\left[Q_{\mathrm{S}}
- \frac{(d-2)(d-4)}{4} + \frac{(d-4)(d-6)}{4}\frac{r^{2}}{l^{2}}\right],
\label{Vscalar}
\eea
where we have introduced the notation $Q_{\mathrm{S}} = k_{\mathrm{S}}^{2}$.
The vector potential is
\bea
V_{\mathrm{V}} = \frac{f}{r^{2}}\left[Q_{\mathrm{V}}
- \frac{(d-2)(d-4)}{4} + \frac{(d-2)(d-4)}{4}\frac{r^{2}}{l^{2}}\right],
\label{Vvector}
\eea
with $Q_{\mathrm{V}} = k_{\mathrm{V}}^{2} -1$.
The tensor potential is
\bea
V_{\mathrm{T}} = \frac{f}{r^{2}}\left[Q_{\mathrm{T}} - \frac{(d-2)(d-4)}{4} + \frac{d(d-2)}{
4}\frac{r^{2}}{l^{2}}\right],
\label{Vtensor}
\eea
with $Q_{\mathrm{T}} = \lambda + 2(d-3)$.

Before solving the above equations, let us first examine the case of a scalar
field $\phi$ of mass $m$ in the background of the massless black hole.  The equation of motion for the scalar
field is
\bea
(\nabla^{2} - m^{2})\phi =0.
\eea
Choosing the ansatz
\bea
\phi = \phi(r)Y(x^{i})e^{\omega t},
\eea
brings the radial equation to the form (\ref{master}), where $\Phi = r^{\frac{d-2}{2}}\phi$.
The potential is given by
\bea
V = \frac{f}{r^{2}}\left[Q
+f^{\prime}\left(\frac{d-2}{2}\right)r +f \frac{(d-2)(d-4)}{4} + m^{2}r^{2}\right],
\label{Vfield1}
\eea
where $\nabla^{2}Y = -QY$.
Since the metric involves the function $f = -1 + \frac{r^{2}}{l^{2}}$, the potential of the scalar field
takes the particularly simple form
\bea
V = \frac{f}{r^{2}}\left[Q
- \frac{(d-2)(d-4)}{4} + \left(\frac{d(d-2)}{4} + m^{2}l^{2}\right)\frac{r^{2}}{l^{2}}\right].
\label{Vfield}
\eea
In \cite{Aros}, this equation was shown to be exactly solvable in terms of hypergeometric
functions.

We now observe that the gravitational potentials (\ref{Vscalar})-(\ref{Vtensor})
have precisely the same structure as the potential of the scalar field (\ref{Vfield}),
for various values of the mass parameter. We have
\bea
{\mathrm{scalar\;\; mode}}:\;\;\;\;\;\; m^{2} l^{2} &=& -2(d-3),\non\\
{\mathrm{ vector\;\; mode}}:\;\;\;\;\;\; m^{2}l^{2} &=& -(d-2),\non\\
{\mathrm {tensor \;\;mode}}:\;\;\;\;\;\; m^{2} l^{2} &=& 0,
\label{masses}
\eea
with the value $Q$ replaced by the appropriate value $Q_{\mathrm{S}}, Q_{\mathrm{V}}, Q_{\mathrm{T}}$.
It should be noted that the simplicity of the potentials in this case is essentially due to the fact
that the mass parameter of the black hole is set to zero.

Our aim now is to solve Eq. (\ref{master}), with potentials (\ref{Vscalar})-(\ref{Vtensor}),
subject to the boundary conditions that $\Phi \rightarrow 0$, at the horizon and at infinity.
To proceed towards the solution of (\ref{master}), we change variables to
\bea
z = 1 - \frac{l^{2}}{r^{2}}.
\eea
Thus, $z=0$ corresponds to the location of the horizon $r = l$, while $z=1$ corresponds to
$r=\infty$.
The master equation then becomes
\bea
z(1-z)\frac{d^{2}\Phi}{dz^{2}} + \left(1 - \frac{3z}{2}\right)\frac{d\Phi}{dz}
+\left[\frac{A}{z} + B + \frac{C}{1-z}\right]\Phi = 0,
\eea
where
\bea
A &=& -\frac{\omega^{2}l^{2}}{4}, \non\\
B &=& \frac{1}{4}\left(\frac{(d-2)(d-4)}{4} - Q\right), \non\\
C &=& -\frac{1}{4}\left(m^{2}l^{2} + \frac{d(d-2)}{4}\right).
\eea
We now define
\bea
\Phi(z) = z^{\alpha}(1-z)^{\beta}F(z).
\eea
The master equation then reduces to the standard form of the hypergeometric equation
\bea
z(1-z)\frac{d^{2}F}{dz^{2}} + [c - (a+b+1)z]\frac{dF}{dz}
- ab F = 0,
\label{hyper}
\eea
provided that
\bea
\alpha &=& \pm\frac{\omega l}{2},\non\\
\beta &=& \frac{1}{4} \pm \frac{1}{4}\sqrt{(d-1)^{2} + 4m^{2}l^{2}},
\label{beta}
\eea
with the coefficients determined as followed
\bea
a &=& \frac{1}{4} + \alpha + \beta  + \frac{1}{4}\sqrt{(d-3)^{2} - 4Q}, \non\\
b &=& \frac{1}{4} + \alpha + \beta - \frac{1}{4}\sqrt{(d-3)^{2} - 4Q},\non\\
c &=& 2 \alpha + 1.
\label{abc}
\eea
Without loss of generality, we can take
\bea
\alpha &=&  \frac{\omega l}{2}, \non\\
\beta &=& \frac{1}{4} - \frac{1}{4}\sqrt{(d-1)^{2} + 4m^{2}l^{2}}.
\label{alphabeta}
\eea

In the neighbourhood of the horizon, the two linearly independent solutions of (\ref{hyper}) are
$F(a,b,c,z)$ and $z^{1-c}F(a-c+1, b-c+1,2-c,z)$. With the choice (\ref{alphabeta}),
the solution which is regular (satisfying Dirichlet boundary conditions) at the horizon is then given by
\bea
\Phi(z) = z^{\alpha}(1-z)^{\beta}F(a,b,c,z).
\eea
Having imposed the Dirichlet boundary condition at the horizon, we can now analytically continue
this solution to infinity. In general, the form of the solution near $z=1$ is given by \cite{Abram}
\bea
\Phi &=& z^{\alpha}(1-z)^{\beta}\frac{\Gamma(c)\Gamma(c-a-b)}{\Gamma(c-a)\Gamma(c-b)}F(a,b,a+b-c+1,1-z)\non\\
&+& z^{\alpha}(1-z)^{\beta + c-a-b}\frac{\Gamma(c)\Gamma(a+b-c)}{\Gamma(a)\Gamma(b)}F(c-a, c-b, c-a-b+1, 1-z).
\label{infty}
\eea
However, special care is needed when $c-a-b$ is an integer. Therefore, we should examine the coefficients closely,
case by case.
As we have seen, the gravitational scalar mode corresponds to a scalar field with mass $m^{2}l^{2} = -2(d-3)$,
the gravitational vector mode corresponds to a scalar field of mass $m^{2} l^{2} = -(d-2)$, and the
gravitational tensor mode
corresponds to a massless scalar field.
It will be useful to record the values of the coefficient $\beta$ given by (\ref{alphabeta}) for the three
gravitational modes, as follows:
\bea
\beta_{S} &=&
\left\{
  \begin{array}{ll}
    0, & d=4, \\
    -\left(\frac{d-6}{4}\right), & d\geq 5,
  \end{array}
\right.\non\\
\beta_{V} &=& -\left(\frac{d-4}{4}\right),\;\; d\geq 4,\non\\
\beta_{T} &=& -\left(\frac{d-2}{4}\right),\;\;d> 4,
\label{beta2}
\eea
where the subscript on $\beta$ specifies the particular mode. From (\ref{abc}), we also note that
$c-a-b = \frac{1}{2} - 2\beta$.

Let us consider first the case in four dimensions. The scalar and vector modes both
have a value of $\beta = 0$, and there is no tensor mode in four dimensions.
Here, $c-a-b = 1/2$, so the analytic continuation to $z=1$ is given by (\ref{infty}). The master field
then takes the form
\bea
\Phi &=& z^{\alpha}\frac{\Gamma(c)\Gamma(c-a-b)}{\Gamma(c-a)\Gamma(c-b)}F(a,b,a+b-c+1,1-z)\non\\
&+& z^{\alpha}(1-z)^{1/2}\frac{\Gamma(c)\Gamma(a+b-c)}{\Gamma(a)\Gamma(b)}F(c-a, c-b, c-a-b+1, 1-z).
\label{4dphi}
\eea
The second term above clearly vanishes at infinity. Furthermore, the perturbation satisfies a Dirichlet boundary condition
at infinity,
$\Phi =0$ at $z=1$, if the coefficients $(a,b,c)$ can be chosen so that the gamma functions in the denominator
of the first term have a pole. Namely, we require
\bea
c-a = -n, \;\; {\mathrm{or}}\;\; c-b = -n,
\label{gammapole}
\eea
where $(n=0,1,2,3,...)$. The existence, or otherwise, of unstable modes then depends on
whether these conditions (\ref{gammapole}) can be implemented. This in turn
depends solely on the eigenvalue of the corresponding Laplacian on the horizon manifold.

For the scalar mode, the coefficients in
the hypergeometric function take the form
\bea
a &=& \frac{\omega l}{2} + \frac{1}{4} + \frac{1}{4}\sqrt{1 - 4 k_{S}^{2}},\non\\
b &=& \frac{\omega l}{2} + \frac{1}{4} - \frac{1}{4}\sqrt{1 - 4 k_{S}^{2}},\non\\
c &=& \omega l + 1,
\eea
where $k_{S}^{2} \geq 0$ is the eigenvalue of the scalar Laplacian on the horizon manifold.
In particular, let us examine the condition $c-a = 0$. This can be written as
\bea
\frac{\omega l}{2} = -\frac{3}{4} + \frac{1}{4}\sqrt{1 - 4 k_{S}^{2}}.
\eea
Thus, in order for an unstable mode with $\omega > 0$ to exist, we require
the scalar Laplacian to have an eigenvalue satisfying  the condition
\bea
k_{S}^{2} < -2.
\label{ks}
\eea
However, $k_{S}^{2} \geq 0$ for arbitrary Einstein horizons \cite{Kodama3}, and thus we conclude that such an unstable scalar mode
does not exist.
Moreover, the constraint $k_{S}^{2}\geq 0$ also ensures that the
conditions $c-a = -n$ for $(n=1,2,3,...)$ and $c-b = -n$ for
$(n=0,1,2,...)$ cannot be satisfied. We conclude that the massless black hole is stable against scalar
perturbations, for arbitrary Einstein horizon.

For the vector modes, we have
\bea
a &=& \frac{\omega l}{2} + \frac{1}{4} + \frac{1}{4}\sqrt{1 - 4(k_{V}^{2} - 1)},\non\\
b &=& \frac{\omega l}{2} + \frac{1}{4} - \frac{1}{4}\sqrt{1 - 4(k_{V}^{2} - 1)},\non\\
c &=& \omega l + 1.
\eea
In this case, the condition $c-a = 0$ requires the vector Laplacian to have an eigenvalue satisfying
\bea
k_{V}^{2} < -1.
\label{kv}
\eea
For general Einstein horizons, $k_{V}^{2} \geq 0$ \cite{Kodama3}, and we conclude that the massless black hole
is stable against vector perturbations.
Collecting the above results, we conclude that (with respect to a choice of
Dirichlet boundary conditions at infinity)
the massless topological black hole is stable in four dimensions, in agreement
with the analysis in \cite{Kodama2}. As one can see from (\ref{4dphi}),
$\Phi$ tends to a constant at infinity, and is therefore
normalizable as it stands. However, as shown in \cite{Wald2},
there is a one-parameter family of self-adjoint extensions of the
operator $A$ in this case.
Imposing a Dirichlet boundary condition at infinity then corresponds
to a particular choice of self-adjoint extension.
The stability properties with respect to the other
self-adjoint extensions
is a problem that warrants further investigation.

Next, we consider all even dimensions greater than four. From (\ref{beta2}), we see that
$\beta \leq 0$ for all perturbations;  furthermore, $c-a-b$ is not an integer.
Thus, the continuation to $z=1$ is given by (\ref{infty})
\bea
\Phi &=& z^{\alpha}(1-z)^{\beta}\frac{\Gamma(c)\Gamma(c-a-b)}{\Gamma(c-a)\Gamma(c-b)}F(a,b,a+b-c+1,1-z)\non\\
&+& z^{\alpha}(1-z)^{\frac{1}{2} - \beta}\frac{\Gamma(c)\Gamma(a+b-c)}{\Gamma(a)\Gamma(b)}F(c-a, c-b, c-a-b+1, 1-z).
\eea
Since $\beta \leq 0$, the second term vanishes automatically at $z=1$.
The Dirichlet boundary condition at infinity can be imposed by choosing the coefficients to satisfy
(\ref{gammapole}). First, we consider the scalar modes. The coefficients are given by
\bea
a &=& \frac{\omega l}{2} - \left(\frac{d-7}{4}\right) + \frac{1}{4}\sqrt{(d-3)^{2} - 4 k_{S}^{2}},\non\\
b &=& \frac{\omega l}{2} - \left(\frac{d-7}{4}\right) - \frac{1}{4}\sqrt{(d-3)^{2} - 4 k_{S}^{2}},\non\\
c &=& \omega l + 1.
\eea
None of the conditions (\ref{gammapole}) can be satisfied unless
\bea
k_{S}^{2} < 0.
\label{kss}
\eea
However, since  $k_{S}^{2} \geq 0$ \cite{Kodama3}, the black hole is stable against scalar perturbations.

For the vector modes, we have
\bea
a &=& \frac{\omega l}{2} - \left(\frac{d-5}{4}\right) + \frac{1}{4}\sqrt{(d-3)^{2} - 4 (k_{V}^{2}-1)},\non\\
b &=& \frac{\omega l}{2} - \left(\frac{d-5}{4}\right) - \frac{1}{4}\sqrt{(d-3)^{2} - 4 (k_{V}^{2} - 1)},\non\\
c &=& \omega l + 1.
\eea
In this case, none of the conditions (\ref{gammapole}) can be satisfied unless
\bea
k_{V}^{2} < -(d-3).
\label{kvv}
\eea
Since $k_{V}^{2} \geq 0$ \cite{Kodama3}, there are no unstable vector perturbations.

For the tensor modes, we have
\bea
a &=& \frac{\omega l}{2} - \left(\frac{d-3}{4}\right) + \frac{1}{4}\sqrt{(d-3)^{2} - 4 [\lambda + 2(d-3)]},\non\\
b &=& \frac{\omega l}{2} - \left(\frac{d-3}{4}\right) - \frac{1}{4}\sqrt{(d-3)^{2} - 4 [\lambda + 2(d-3)]},\non\\
c &=& \omega l + 1.
\eea
In this case, the condition $c-a = 0$ can be satisfied if the Lichnerowicz operator has an eigenvalue
\bea
\lambda < -4(d-2).
\label{Lich}
\eea
This result was obtained previously in \cite{Gibbons}. Furthermore, if $\lambda \geq -4(d-2)$, then we
conclude that the black hole is stable against tensor perturbations.
For the scalar perturbation in six dimensions, we notice
from (\ref{beta2}) that $\beta_{S} = 0$, and thus the perturbation is
normalizable as it stands.
Thus, the choice of Dirichlet boundary conditions
at infinity corresponds to a particular choice of
self-adjoint extension \cite{Wald2}.

Turning now to odd dimensions, let us first consider the case of $d=5$. The subtlety here is that
$c-a-b$ is an integer, and the analytic continuation to $z=1$ contains logarithmically divergent terms.
For the scalar perturbation, we have $\beta_{S} = 1/4$ and $c-a-b = 0$. The master field
near $z=1$ is then given by \cite{Abram}
\bea
\Phi &=& z^{\alpha}(1-z)^{1/4} \frac{\Gamma(a+b)}{\Gamma(a)\Gamma(b)}\sum_{n=0}^{\infty}
\frac{(a)_{n}(b)_{n}}{(n!)^{2}}[2\psi(n+1)
- \psi(a+n) \non\\
&-&\psi(b+n) - {\mathrm{ln}}(1-z)](1-z)^{n},\non\\
\eea
where $(a)_{n} = \Gamma(a+n)/\Gamma(a)$, and $\psi(z) = \Gamma^{\prime}(z)/\Gamma(z)$.
In this case, $\Phi$ automatically vanishes at infinity.
However, as shown in \cite{Wald2}, there is a one-parameter family of
self-adjoint extensions of the corresponding operator $A$ for pure
anti-de Sitter space. In order to proceed, we must then
make a choice of self-adjoint extension, and thus make a
corresponding choice of boundary conditions
at infinity. We shall require $(1-z)^{-1/4}\Phi$ to vanish at infinity.
This is achieved by choosing $a = -n$ or $b=-n$. However, since,
$c-a-b = 0$, this can be re-written as (\ref{gammapole}),
namely $c-a = -n$ or $c-b = -n$.
We have already seen that these conditions lead to
the constraint (\ref{kss}), with the conclusion
that no such unstable modes exist.
The above choice of boundary condition corresponds to what is
termed generalized Dirichlet-Neumann in \cite{Wald2}.

For the vector modes, we have $\beta_{V} = -1/4$, and $c-a-b = 1$.
The master field is then given by
\bea
\Phi =&=& z^{\alpha}(1-z)^{-1/4}F(a,b,a+b+1,z),
\eea
where, for $(m=1,2,3,...)$, we have
\bea
F(a,b,a+b+m,z) &=& \frac{\Gamma(m)\Gamma(a+b+m)}{\Gamma(a+m)\Gamma(b+m)}\sum_{n=0}^{m-1}
\frac{(a)_{n}(b)_{n}}{n!(1-m)_{n}}(1-z)^{n}\non\\
&-&
\frac{\Gamma(a+b+m)}{\Gamma(a)\Gamma(b)}(z-1)^{m}\sum_{n=0}^{\infty}\frac{(a+m)_{n}(b+m)_{n}}{n!(n+m)!}
(1-z)^{n}[{\mathrm{ln}}(1-z) \non\\
&-& \psi(n+1) -\psi(n+m+1) + \psi(a+n+m) + \psi(b+n+m)].
\label{5d}
\eea
The conditions for the existence of unstable vector modes are then given by $a+1 = -n$ or $b+1 = -n$. Note that these conditions
also ensure the vanishing of the logarithmic terms in (\ref{5d}). Since $c-a-b=1$, we can equivalently write
these conditions as $c-a = -n$ or $c-b = -n$, which have already been treated in (\ref{kvv}).
Again, we conclude that no unstable vector modes exist.

Finally, the tensor modes have $\beta_{\mathrm{T}} = -3/4$, with $c-a-b = 2$. The Dirichlet
boundary condition at infinity is enforced by
setting $a+2 = -n$ or $b+2 =-n$. Since this is equivalent to $c-a = -n$ or $c-b = -n$, we recover the constraint
(\ref{Lich}).
The generalization to all odd dimensions follows suit, with the knowledge that $\beta < 0$
for all perturbations, and $c-a-b = m$, with $(m = 1,2,3,...)$.
Thus, the criteria for the existence of unstable modes can again be written in the form (\ref{gammapole}), and the conclusions
are as in the even-dimensional case.

We can now collect these results to establish a necessary and sufficient condition for stability
of the massless topological black hole. As we have seen, the massless black hole is stable against
scalar and vector perturbations in all dimensions, for an arbitrary Einstein horizon. The non-trivial constraint
on stability arises from the tensor modes.
We can state that a necessary
and sufficient condition for
stability of the black hole is given by,
\bea
\lambda \geq -4(d-2) \Leftrightarrow {The \;\; Massless\;\;Black\;\; Hole\;\; is\;\;
Stable}.
\label{theorem1}
\eea
In \cite{Gibbons}, the tensor perturbations alone were analyzed, and the condition
$\lambda < -4(d-2)$ was thus obtained as a sufficient condition for instability of the black hole.
By obtaining the explicit solution for the scalar and vector
modes, we have elevated
the result of \cite{Gibbons} to a necessary and sufficient condition for stability.

If we take the horizon manifold to have constant curvature, then it is given by hyperbolic space $H^{d-2}$, or
a quotient $H^{d-2}/\Gamma$, where $\Gamma$ is a suitable discrete subgroup of the isometry group of hyperbolic
space.
For such manifolds, the spectrum of the Lichnerowicz operator is bounded below by $\lambda \geq -2(d-2)$ \cite{Gibbons},
which satisfies the condition of (\ref{theorem1}).
We conclude that this class of massless topological black holes is stable. A second class of stable black holes
is provided by taking the horizon to be a
negative scalar curvature Einstein-K\"{a}hler manifold. In this case, the spectrum of the
Lichnerowicz operator is bounded by $\lambda \geq -2(d-3)$ \cite{Gibbons, Friedan}, which again satisfies the condition (\ref{theorem1}).
For general Einstein horizons, we have reduced the stability issue to a simple bound on the Lichnerowicz spectrum.

\section{Stability of Black Holes with $M<0$}

Explicit solutions to the perturbation equations for black holes of non-zero mass are not available.
In order to study the stability properties of such black holes, we
appeal to general arguments based on positivity of the corresponding
potentials in the perturbations equations. The requirement of
positivity of a particular potential (scalar, vector, tensor), then provides a sufficient
condition for stability of the black hole with respect to the corresponding
scalar, vector, or tensor perturbation. For positive mass black holes $M>0$ in dimensions $d>4$, no conclusion
regarding stability against scalar perturbations can be made from such positivity arguments \cite{Kodama3}-\cite{Kodama5}.
However, as we have seen, there is a class of black holes with negative curvature horizon, for which the
mass parameter $M$
can take a range of negative values, namely, $M_{\mathrm{crit}} \leq M < 0$.

We recall that the basic Schr\"{o}dinger equation is of the form (\ref{Aoperator1}).
In order to establish stability of the black hole, we need to
prove that $A$ can be extended to a positive definite self-adjoint operator.
The expectation value of $A$ is given by \cite{Kodama2}
\bea
(\Phi, A\Phi) = -\left[\Phi^{*}\frac{d\Phi}{dr_{*}}\right]_{Boundary}
+\int dr_{*}\left(\mid\!\frac{d\Phi}{dr_{*}}\!\mid^{2} + V \mid\!\Phi\!\mid^{2}\right).
\eea
Thus, if boundary conditions are imposed which render the boundary term zero,
and if $V\geq 0$, then we conclude that $A$ is positive definite.
A more powerful approach is to use an $S$-deformation of the potential $V$ in the following way \cite{Kodama2}.
One defines a derivative operator
\bea
\tilde{D} = f\frac{d}{dr} + S,
\eea
where $S$ is some function of $r$. Then, we can write the expectation value in the form
\bea (\Phi,A\Phi) = -[\Phi^{*}\tilde{D}\Phi]_{Boundary} + \int\;
dr_{*}(\mid\! \tilde{D}\Phi\!\mid^{2} + \tilde{V}\!\mid\!\Phi\!\mid^{2}),
\label{expectation}
\eea
where the deformed potential is now
\bea
\tilde{V} = V + f\frac{dS}{dr} - S^{2}.
\eea
The boundary term in (\ref{expectation}) vanishes when Dirichlet
boundary conditions on $\Phi$ are chosen. Thus,
if Dirichlet boundary conditions
are chosen, and if $\tilde{V} \geq 0$,
we can conclude that $A$ is a positive definite operator.
The asymptotic (large $r$) form of the negative mass black hole metric
coincides with the metric of the massless black hole.
Thus, we must indeed impose Dirichlet boundary conditions at infinity
for all perturbations, except for certain perturbations in dimensions
four, five, and six.
In the following analysis, we shall adopt a choice of
Dirichlet boundary conditions in these cases also.

Let us first consider the constraints that arise by demanding positivity of the potentials
$V$ and $\tilde{V}$ for vector modes.
From (\ref{Vector}), we note that the potential can be written in the form
\bea
V_{V}(r) = \frac{f}{r^{2}} \left[k_{V}^{2} -1 + \frac{(d-2)(d-4)}{4}f
- (d-2)(d-1)\frac{\omega_{d}M}{2r^{d-3}}\right].
\eea
For negative mass black holes, we clearly have a positive potential if
\bea
k_{V}^{2} \geq 1.
\eea
However, in this case, we can achieve a
better bound by using the $S$-deformation technique, with \cite{Kodama2}
\bea
S = \frac{(d-2)f}{2r}.
\eea
The deformed potential is then given by
\bea
\tilde{V}_{V}(r) = \frac{f}{r^{2}}[k_{V}^{2} + (d-3)].
\label{deformedvector}
\eea
Thus, positivity of the deformed potential is guaranteed when
\bea
k_{V}^{2} \geq -(d-3).
\label{vectorbound}
\eea
Since the eigenvalues of the vector Laplacian on a generic Einstein space satisfy the condition $k_{V}^{2} \geq 0$,
we have established that this bound is satisfied. Thus,
$\tilde{V}_{V} \geq 0$. Hence, negative mass black holes are stable against vector perturbations for arbitrary
Einstein horizons.

Moving on to the tensor modes, we note that the potential takes the form
\bea
V_{T}(r) = \frac{f}{r^{2}} \left[\lambda +[3(d-2) - 2] + \frac{(d-2)d}{4}f
+ (d-2)(d-1)\frac{\omega_{d}M}{2r^{d-3}}\right].
\label{Tensor2}
\eea
Since $M < 0$, the last term in the potential is negative. However, the most
negative value that it can take is attained when
\bea
\frac{\omega_{d}M}{r^{d-3}} = \frac{\omega_{d}M_{\mathrm{crit}}}{r_{\mathrm{crit}}^{d-3}} = -\frac{2}{d-1}.
\label{Mcrit}
\eea
Inserting this value into (\ref{Tensor2}), we find that the potential is positive if the Lichnerowicz spectrum satisfies
the bound
\bea
\lambda \geq -2(d-3).
\label{tensorbound}
\eea
The $S$-deformed potential is this case can be obtained by
taking \cite{Kodama2}
\bea
S = -\frac{(d-2)f}{2r}.
\eea
This leads to a deformed potential of the form
\bea
\tilde{V}_{T}(r) = \frac{f}{r^{2}}[\lambda + 2(d-3)].
\eea
Positivity of $\tilde{V}_{T}$ then requires the same bound as in (\ref{tensorbound}).

The analysis of the scalar mode is the most lengthy.
In this case, we appeal only to the
$S$-deformation of the potential, which can be achieved by
choosing \cite{Kodama3}
\bea
S = \frac{f}{r^{\frac{d-2}{2} -1}H}\frac{d}{dr}(r^{\frac{d-2}{2} -1}H).
\eea
After a lengthy calculation, one finds the deformed potential to be
\bea
\tilde{V}_{S}(r) = \frac{k_{S}^{2}f}{2r^{2}H}[2\mu - (d-1)(d-4)x],
\eea
where
\bea
H = k_{S}^{2} + (d-2) +\frac{(d-2)(d-1)}{2}x.
\eea
The key point now is that we are considering negative mass black holes, and thus
$x = \omega_{d}M/r^{d-3} <0$. However, once again, the most negative that $x$ can become is given
by (\ref{Mcrit}), and hence the value of $H$ is at least $k_{S}^{2}$. The eigenvalues of the scalar Laplacian
on a generic Einstein manifold satisfy the condition $k_{S}^{2} \geq 0$.
Thus, for negative mass black holes, the deformed scalar potential is positive definite for generic Einstein horizons.

Collecting these results, we can present a sufficient condition for the stability of negative mass black holes
with respect to all perturbations, namely
\bea
\lambda \geq -2(d-3) \Rightarrow The\;\; M<0\;\; Black\;\; Hole\;\; is\;\; Stable.
\eea
We recall that the Lichnerowicz spectrum for negative scalar curvature Einstein-K\"{a}hler manifolds satisfies
the bound $\lambda \geq -2(d-3)$. We thus conclude that all negative mass black holes with an horizon of this type are stable.

Incidentally, it is also useful to consider the constraints which arise form positivity of the potentials
in the massless case. For vector modes, the deformed potential (\ref{deformedvector}) again leads to the constraint
(\ref{vectorbound}). Thus, vector modes
are stable in all dimensions. For tensor modes, positivity of the potential (\ref{Tensor2}) with $M=0$ actually
leads to a stronger constraint that the deformed potential, namely $\lambda \geq -3(d-2) +2$.
For scalar modes, we take $S = (d-4)\frac{f}{2r}$, leading to a deformed potential $\tilde{V}_{S} = \frac{k_{S}^{2}f}{r^{2}}$.
Thus, scalar modes are stable in all dimensions. These results are consistent with the necessary and sufficient
condition for stability that we derived earlier, based on the explicit solution of the equations.

\section{Discussion}
A complete investigation of the classical stability properties of all higher-dimensional black holes is an important problem.
Until quite recently, the stability of the asymptotically flat Schwarzschild black hole in all dimensions was an open issue.
However, following the analysis of tensor perturbations in \cite{Gibbons}, a gauge invariant formalism
for all gravitational perturbations was established by Ishibashi and Kodama \cite{Kodama1}.
Armed with this elegant and powerful formalism, the Schwarzschild black
hole was indeed shown to be stable
in all dimensions \cite{Gibbons, Kodama2}.

The purpose of the present paper has been to extend this analysis to other classes of black holes.
We have focused, in particular, on topological black holes in anti-de Sitter space, for which the horizon is a negative
curvature Einstein manifold. A special feature in this case is the presence of zero mass and negative mass black holes.
For the zero mass black hole, we showed that the equations for scalar, vector, and tensor perturbations assumed
a unified form, which was exactly solvable in terms of hypergeometric functions.
We showed that the massless black hole
is stable against scalar and vector perturbations in all dimensions.
The only dangerous mode is therefore the tensor mode. However,
using the exact solutions, we succeeded in deriving a necessary and sufficient condition for stability in all dimensions.
This condition was expressed solely in terms of the spectrum of the Lichnerowicz operator on the horizon
manifold.
The analysis presented here thus elevated the results of \cite{Gibbons} on tensor perturbations
to a necessary and sufficient condition
for stability.
While the form of the Lichnerowicz spectrum on general Einstein spaces is not known, there are some examples
where enough information is available to establish stability. In particular, we proved
that massless black holes with constant curvature (hyperbolic) horizons, or Einstein-K\"{a}hler horizons, are stable
in all dimensions.

For negative mass black holes, an explicit solution to the perturbation equations is not available.
Nevertheless, using the requirement of positivity of the gravitational potentials, along with the $S$-deformation technique,
a sufficient condition for stability for all negative mass black holes was derived.
As in the massless case,
we showed that these black holes are stable again all scalar and vector perturbations, the only dangerous mode being
the tensor mode.
Again, the sufficient condition for stability is expressed in terms of the Lichnerowicz spectrum.
In particular, we concluded that negative mass black holes with Einstein-K\"{a}hler horizons are stable in all dimensions.

In general, the choice of boundary conditions appropriate to the
stability problem are determined by requiring normalizability of
the perturbation.
In dimensions $d\geq 7$, this leads to the requirement of
Dirichlet boundary conditions both at the horizon and infinity,
However, for certain perturbations in dimensions four, five, and six,
there is a subtlety in the choice of boundary conditions at infinity.
This follows from the fact that there is a one-parameter family
of self-adjoint extensions of the
perturbation operator in these cases \cite{Wald2},
with a corresponding freedom in the choice of boundary conditions.
In these special cases, we considered the stability issue with a choice
of Dirichlet boundary condition. The stability properties with
respect to the other possible choices warrants further investigation.


\end{document}